\documentclass[12pt]{article}
\usepackage{graphicx}

\newcommand\pubnumber{CIPANP2015-Ito}
\newcommand\pubdate{\today}

\newcommand{\pienu}{${\pi}^+{\rightarrow}e^+{\nu}_e$}
\newcommand{\pienug}{${\pi}^+{\rightarrow}e^+{\nu}_e{\gamma}$}
\newcommand{\pimunu}{${\pi}^+{\rightarrow}{\mu}^+{\nu}_{\mu}$}
\newcommand{\pimunug}{${\pi}^+{\rightarrow}{\mu}^+{\nu}_{\mu}{\gamma}$}
\newcommand{\pimue}{${\pi}^+{\rightarrow}{\mu}^+{\rightarrow}e^+$}
\newcommand{\mue}{${\mu}^+{\rightarrow}e^+{\nu}_e\overline{{\nu}}_{\mu}$}

\def\Title#1{\begin{center} {\Large #1 } \end{center}}
\def\Author#1{\begin{center}{ \sc #1} \end{center}}
\def\Address#1{\begin{center}{ \it #1} \end{center}}

\newcommand\pubblock{\rightline{\begin{tabular}{l} \pubnumber\\
         \pubdate  \end{tabular}}}
\newenvironment{Abstract}{\begin{quotation}  }{\end{quotation}}
\newenvironment{Presented}{\begin{quotation} \begin{center} 
             PRESENTED AT\end{center}\bigskip 
      \begin{center}\begin{large}}{\end{large}\end{center} \end{quotation}}
\def\Acknowledgements{\bigskip  \bigskip \begin{center} \begin{large}
             \bf ACKNOWLEDGEMENTS \end{large}\end{center}}

\begin{document}
\begin{titlepage}
\pubblock

\vfill
\Title{Status of the TRIUMF PIENU Experiment}
\vfill
\Author{ \underline{S.Ito}$^1$, A.Aguilar-Arevalo$^2$, M.Aoki$^1$, M.Blecher$^3$, D.I.Britton$^4$, D.A.Bryman$^5$, D. vom Bruch$^5$, S.Chen$^6$, J.Comfort$^7$, S.Cuen-Rochin$^5$, L.Doria$^8$, P.Gumplinger$^8$, A.Hussein$^9$, Y.Igarashi$^a$, S.Kettell$^b$, L.Kurchaninov$^8$, L.Littenberg$^b$, C.Malbrunot$^{c,5}$, R.E.Mischke$^8$, T.Numao$^8$, D.Protopopescu$^4$, A.Sher$^8$, T.Sullivan$^5$, and D.Vavilov$^8$}
\Address{$^1$Osaka University, Toyonaka, Osaka, 560-0043, Japan\\
$^2$Instituto de Ciencias Nucleares, Universidad Nacional Aut$\acute{\rm o}$noma de Mexico, M$\acute{\rm e}$xico\\
$^3$Virginia Tech., Blacksburg, VA 24061, USA\\
$^4$University of Glasgow, Glasgow, UK\\
$^5$University of British Columbia, Vancouver, B.C. V6T 1Z1, Canada\\
$^6$Tsinghua University, Beijing, 100084, China\\
$^7$Arizona State University, Tempe, AZ 85287, USA\\
$^8$TRIUMF, 4004 Wesbrook Mall, Vancouver, B.C. V6T 2A3, Canada\\
$^9$University of Northern British Columbia, Prince George, B.C. V2N 4Z9, Canada\\
$^a$KEK, 1-1 Oho, Tsukuba-shi, Ibaraki, 305-0801, Japan\\
$^b$Brookhaven National Laboratory, Upton, NY 11973-5000, USA\\
$^c$Present address: CERN, Gen$\acute{\rm e}$ve 23 CH-1211, Switzerland}
\vfill
\begin{Abstract}
The PIENU experiment at TRIUMF aims to measure the pion decay branching ratio $R={\Gamma}({\pi}^+{\rightarrow}e^+{\nu}_e({\gamma}))/{\Gamma}({\pi}^+{\rightarrow}{\mu}^+{\nu}_{\mu}({\gamma}))$ with precision $<0.1$\% to provide a sensitive test of electron-muon universality in weak interactions. 
The current status of the PIENU experiment is presented. 
\end{Abstract}
\vfill
\begin{Presented}
Twelfth Conference on the Intersections of Particle and Nuclear Physics (CIPANP2015)\\
Vail, Colorado, USA, May 19--24, 2015
\end{Presented}
\vfill
\end{titlepage}
\def\thefootnote{\fnsymbol{footnote}}
\setcounter{footnote}{0}

\section{Introduction}
The charged pion decay branching ratio $R={\Gamma}$({\pienu} + {\pienug})/${\Gamma}$({\pimunu} + {\pimunug}) is one of the most precisely calculated observables in the Standard Model (SM) involving quarks \cite{ANNREV}. 
The most recent theoretical evaluation \cite{ANNREV}, \cite{Theory} gives 
\begin{eqnarray}
R_{SM}=(1.2352{\pm}0.0002){\times}10^{-4}. 
\end{eqnarray}
Precise measurement of $R$ provides one of the most stringent tests of the hypothesis of electron-muon universality in weak interactions. 
The previous experimental values of the branching ratio are
\begin{eqnarray}
R_{EXP1}&=&[1.2265{\pm}0.0034(stat){\pm}0.0044(syst)]{\times}10^{-4} ({\rm TRIUMF},~1992)~\cite{E248},
\end{eqnarray}
and
\begin{eqnarray}
R_{EXP2}&=&[1.2346{\pm}0.0035(stat){\pm}0.0036(syst)]{\times}10^{-4} ({\rm PSI},~1993)~\cite{PSI}
\end{eqnarray}
indicating that there is a room for improvement by two orders of magnitude in precision. 

The goal of the PIENU experiment at TRIUMF is to improve the accuracy of the branching ratio measurement by a factor of 5, to $<$0.1\% resulting in 0.05\% precision in the universality test. 
This precision also allows potential access to new physics up to the mass scale of 1000 TeV for helicity unsuppressed pseudoscalar interactions \cite{ANNREV}. 
Examples of the new physics probed include R-parity violating SUSY \cite{SUSY}, heavy neutrino mixing \cite{Neutrino}, excited gauge bosons, leptoquarks \cite{Leptoquarks}, compositeness, and the effects of charged Higgs bosons.
In the following, the result of the analysis of an initial data set is presented \cite{PRL}. 

\section{The PIENU Experiment}
The PIENU experiment was located at the TRIUMF M13 beam line \cite{M13}. 
The positive charged beam momentum was $P$=75${\pm}1$ MeV/{\it c} with a rate about 70 kHz, and composition 84\% ${\pi}^+$, 14\% ${\mu}^+$, and 2\% $e^+$.

\begin{figure}[]
\centering
\includegraphics[width=11cm,clip]{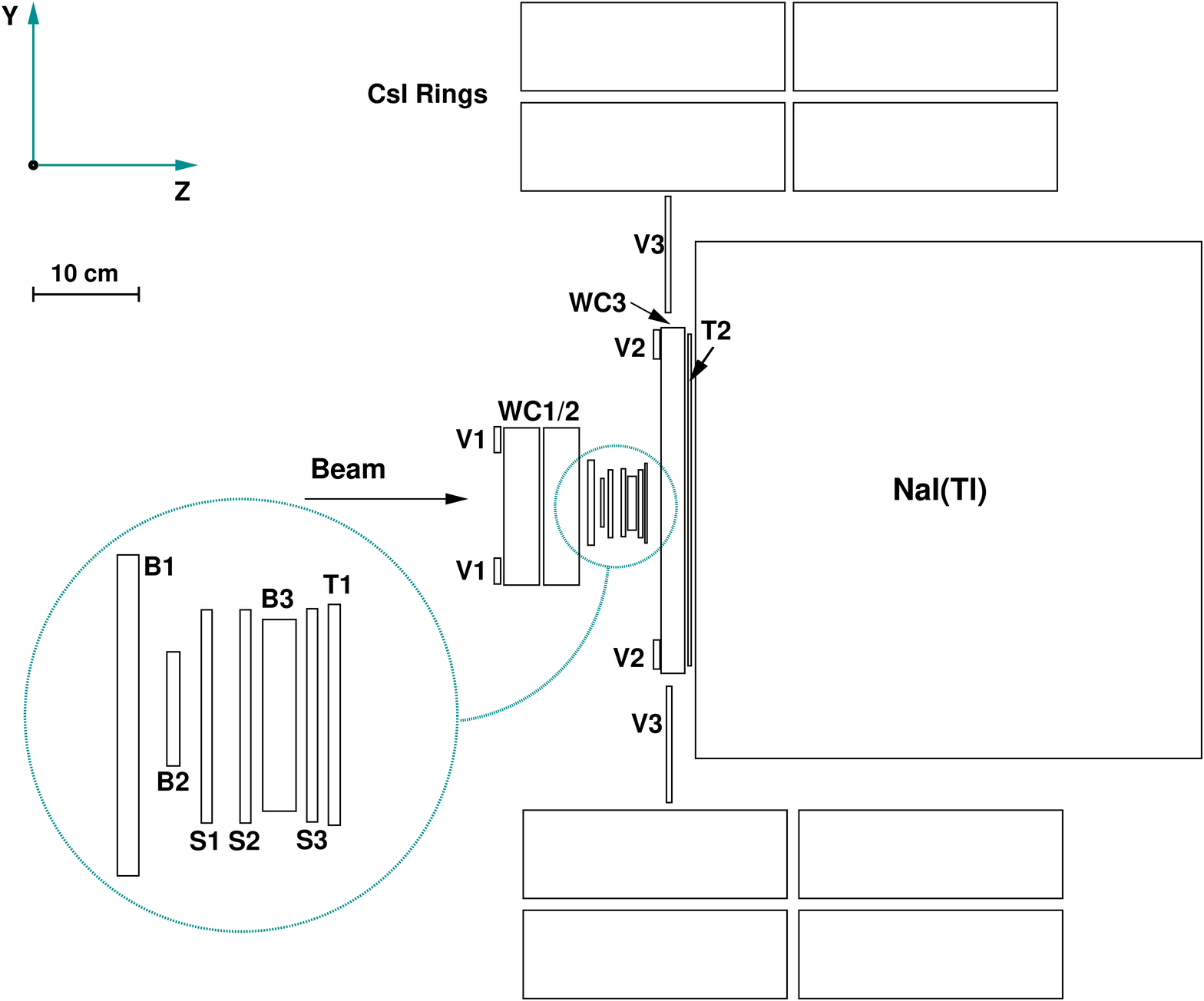}
\caption{Schematic of the PIENU detector.}
\label{fig:concept_nim2}
\end{figure}

Figure \ref{fig:concept_nim2} shows the schematic of the PIENU detector. 
Pion beam tracking was provided by two three-plane wire chambers (WC1 and WC2) located at the exit of the beam line. 
Following WC2, the beam was degraded by two thin plastic scintillators B1 and B2 (beam counters) used for time and energy loss measurements in order to identify the beam pions. 
The beam counters were followed by two sets of Si strip detectors (S1 and S2). 
The pions stopped in the center of an 8 mm plastic scintillator target (B3) and decayed at rest. 

In order to reconstruct the tracks and define the acceptance of decay positrons, another Si strip detector (S3) and three layers of wire chamber (WC3) were employed. 
Two thin plastic scintillators (telescope counters T1 and T2) were used to measure decay times and to define the on-line trigger. 
Triggered decay positrons entered a large single NaI(Tl) crystal calorimeter (48 cm diameter ${\times}$ 48 cm width). 
Two layers of pure CsI crystals surrounded the NaI(Tl) crystal for shower leakage detection. 
Three veto scintillators (V1$-$3) were installed to cover inactive material. 
The solid angle acceptance was 20\%, about 10 times larger than in the previous TRIUMF experiment \cite{E248}.
Details of the PIENU detector are described in reference \cite{NIMA}.

\section{Analysis}
\begin{figure}[]
\centering
\includegraphics[width=11cm,clip]{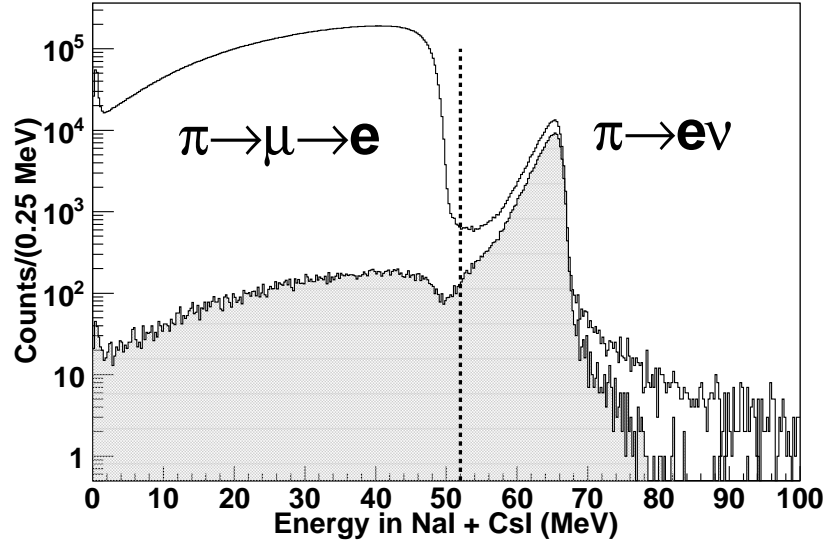}
\caption{Energy spectra of positrons in the time region 5$-$35 ns without and with (shaded) {\pimue} suppression cuts (see the text). The vertical line indicates the $E_{Cut}=52$ MeV.}
\label{fig:BinaCsI}
\end{figure}

\begin{figure}[]
\centering
\includegraphics[width=15.2cm,clip]{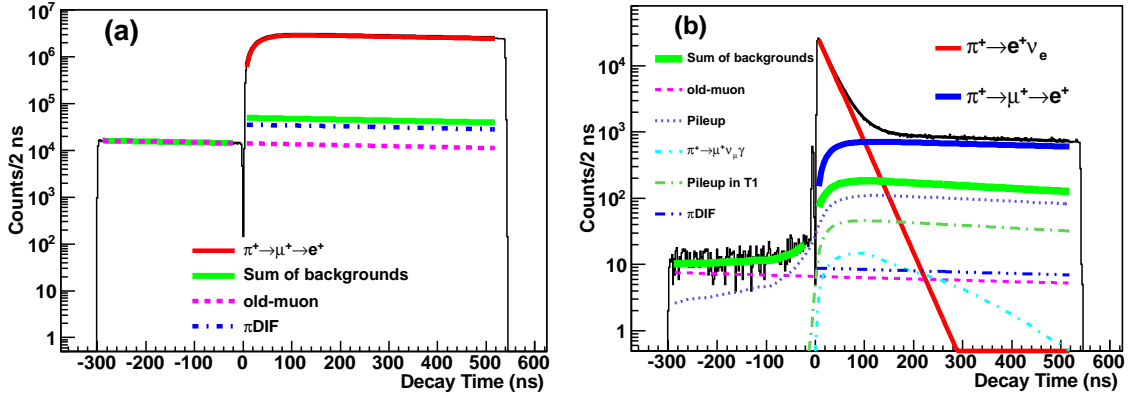}
\caption{Time spectra in the low (a) and high (b) energy regions. Horizontal axes are decay time. The solid red lines in the low and high energy regions are due to {\pimue} and {\pienu} respectively. Other lines indicate background components and the solid green lines in both regions are the sums of the background components (see text for details).}
\label{fig:TS}
\end{figure}

The raw branching ratio was obtained from the ratio of positron yields from {\pienu} decays ($E_e=69.8$ MeV) and {\pimunu} decays followed by $\mu^+\rightarrow e^+ \nu_e \overline\nu_{\mu}$ decays (the {\pimue} decay chain, $E_e=0.5-52.8$ MeV). 
Figure \ref{fig:BinaCsI} shows the energy spectrum of decay positrons in the region 5$-$35 ns obtained by the NaI(Tl) and CsI. 
Decay positrons were divided into two energy regions separated at $E_{Cut}=52$ MeV. 
The time spectra of decay positrons in the low and high energy regions are shown in Figure \ref{fig:TS}. 
Those spectra were fitted to functions including background components simultaneously to extract the raw branching ratio. 

In the low energy time spectrum, the main components were {\pimue} decays at rest ($\frac{{\lambda}_{\pi}{\lambda}_{\mu}}{{\lambda}_{\pi}-{\lambda}_{\mu}}(e^{-{\lambda}_{\mu}t}-e^{-{\lambda}_{\pi}t})$ starting at $t=0$), $\pi$ decay-in-flight ($\pi$DIF) events upstream of B3 (${\lambda}_{\mu}e^{-{\lambda}_{\mu}t}$ starting at $t=0$), and previously stopped muons (``old-muon'', ${\lambda}_{\mu}e^{-{\lambda}_{\mu}t}$). 

The main component in the high energy time spectrum was the {\pienu} decays (${\lambda}_{\pi}e^{-{\lambda}_{\pi}t}$). 
The major backgrounds were from muon decays ({\pimue}, $\pi$DIF, and old-muons). 
These components have the same time distributions as in the low energy region. 
{\pimue}, $\pi$DIF, and old-muon decays in the high energy region are shown as the solid blue line, the dashed dark blue line, and the dashed pink line in Figure \ref{fig:TS}(b) (colors on-line). 

Another background in the high energy region came from radiative pion decays {\pimunug} (branching ratio, $2{\times}10^{-4}$ \cite{pimunug}) followed by {\mue} decays. 
In this case, the time of the $\gamma$-ray is different from that of the decay positron. 
These contributions were estimated by Monte Carlo (MC) simulation using waveform templates for the NaI(Tl) and CsI detectors. 
This distribution is shown as the light blue line in Figure \ref{fig:TS}(b).

The distribution shown by the dashed violet line in the high energy region in Figure \ref{fig:TS}(b) is the pileup component of {\pimue} decays plus old-muon decays. The shape of this background was obtained by MC simulation using the pulse shapes of the NaI(Tl) and CsI detectors.

The pileup cut was based on the pulse shape in T1. 
However, events with two T1 hits within the double pulse resolution of T1 (${\sim}15$ ns) were accepted. 
The amplitude was estimated by artificially increasing the double pulse resolution up to 200 ns. 
This distribution is shown by the dashed green line in Figure \ref{fig:TS}(b). 

To reduce possible bias, the raw branching ratio was shifted (``blinded'') by a hidden random value within 1\% during the initial analysis procedure. 
Prior to unblinding, all cuts and corrections were determined and the stability of the result against variations of each cut was reflected in the systematic uncertainty estimate. 

\section{Systematic Corrections}

\subsection{Tail Correction}
The largest correction to the branching ratio came from knowledge of the low energy {\pienu} tail events below $E_{Cut}$. 
In order to evaluate the amount of {\pienu} tail, {\pimue} events were suppressed using an early decay-time region 5$-$35 ns (unfilled histogram in Figure \ref{fig:BinaCsI}), pulse shape and the total energy in B3, and beam tracking information. 
The resulting {\pimue} suppressed positron energy spectrum is shown by the shaded histogram in Figure \ref{fig:BinaCsI}. 
In order to evaluate the correction for the {\pienu} low energy tail, the remaining {\pimue} events of the suppressed spectrum were subtracted and the {\pienu} tail was obtained. 
Additionally, since the total energy cut in B3 removed {\pienu} events with Bhabha scattering and larger energy loss in B3, MC simulation was used to take account of those events. 
These procedures were based on the method of the previous TRIUMF experiment \cite{E248} and provided a lower bound on this correction. 

\begin{figure}[]
\centering
\includegraphics[width=11cm,clip,trim=2cm 0cm 2cm 0cm]{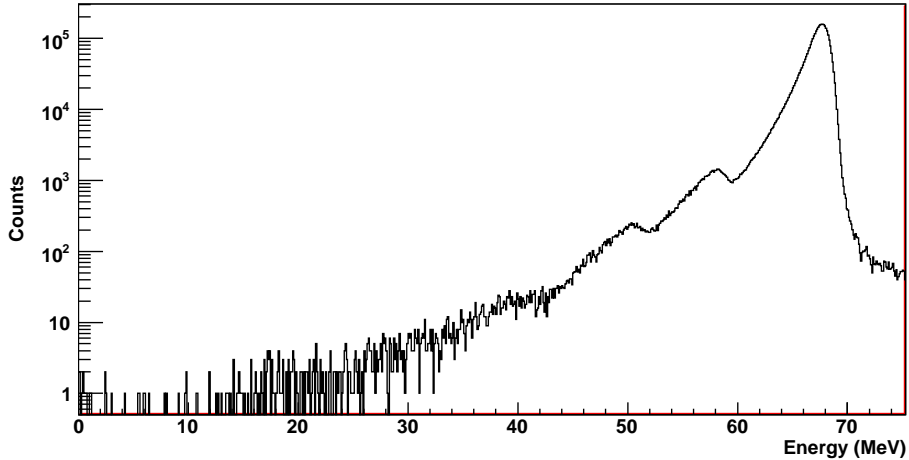}
\caption{Tail measurement at 0 degrees. Below the main peak at 68 MeV, the three bumps were due to photo-nuclear absorption in the NaI(Tl) calorimeter \cite{NIM}.}
\label{fig:LineShape}
\end{figure}

In order to empirically determine an upper bound of the {\pienu} low energy tail, a special data set was taken using mono-energetic positron beams. 
The crystal calorimeter was rotated to obtain different angles of entry. 
Figure \ref{fig:LineShape} shows the energy spectrum at 0 degrees. 
It shows the peak of the beam positrons as well as three lower energy bumps due to photo-nuclear absorption followed by neutron escape from the NaI(Tl) crystal \cite{NIM}.

Combining the lower and upper bounds, the {\pienu} low energy tail was estimated to be $3.07{\pm}0.12$\%.

\subsection{Acceptance Correction}
Energy dependent effects changed the relative acceptance of {\pienu} and {\pimue} events. 
The acceptance correction relied on MC calculations including multiple Coulomb scattering, Bhabha scattering, positron annihilation-in-flight, and trigger losses. 
The ratio of acceptance of {\pienu} and {\pimue} was estimated within the uncertainty of 0.03\%. 
The uncertainties of the detector geometry and the pion beam stopping position were also included in the error estimate.

\subsection{Other Corrections}
For the decays-in-flight of muons from {\pimunu} in B3 ($\mu$DIF), Lorentz boosting raises the positron energy. 
Since $\mu$DIF had the same time distribution as {\pienu} decays, $\mu$DIF events inflated the apparent number of {\pienu} decays. 
The contribution of $\mu$DIF was estimated by MC. 

Possible energy-dependent effects on T1 were studied using decay positrons from muons stopped at the center of B3. 
The effects of the uncertainties of the muon and pion lifetimes \cite{LMLP} were also included.

\section{Status and Conclusion}
Data taking of the PIENU experiment was completed in 2012, and the analysis of a partial data set taken in 2010 has been completed \cite{PRL}. 
The partial data set corresponds to $4{\times}10^5$ {\pienu} decay events. 
For this data set with all corrections 
\begin{eqnarray}
R_{EXP}&=&[1.2344{\pm}0.0023(stat){\pm}0.0019(syst)]{\times}10^{-4}
\end{eqnarray}
which is consistent with previous work and the SM prediction. 
The present result improves the test of electron-muon universality compared to previous experiments by a factor of two: $g_e/g_{\mu}=0.9996{\pm}0.0012$ for the charged current. 

The analysis of the remaining data is in progress. 
The statistical uncertainty will be improved by a factor of 3 since the full data set corresponds to 10 times higher statistics. 
It is anticipated that the systematic uncertainty will also be substantially improved.

\Acknowledgements
This work was supported by the Natural Sciences and Engineering Research Council Grant SAPPJ 157985-2013 and TRIUMF through a contribution from the National Research Council of Canada, and by the Research Fund for the Doctoral Program of Higher Education of China, by CONACYT doctoral fellowship from Mexico, and by JSPS KAKENHI Grants No. 18540274, No. 21340059, and No. 24224006 in Japan.  We are grateful to Brookhaven National Laboratory for the loan of the crystals and to the TRIUMF operations, detector, electronics, and DAQ groups for their engineering and technical support.


\begin{thebibliography}{99}


\bibitem{ANNREV}D. Bryman, W. Marciano, R. Tschirhart and T. Yamanaka,  Ann. Rev. Nucl. Part. Sci. {\bf 61}, 331 (2011).

\bibitem{Theory}V. Cirigliano and I. Rosell, {\it  JHEP} {\bf 0710}, 005 (2007).

\bibitem{E248}D.I. Britton {\it et al}., {\it Phys. Rev. Lett}. {\bf 68}, 3000 (1992). and D.I. Britton {\it et al}., {\it Phys. Rev}. {\bf D49}, 28 (1994).

\bibitem{PSI}G. Czapek {\it et al}., {\it Phys. Rev. Lett}. {\bf 70}, 17 (1993).

\bibitem{SUSY}M.J. Ramsey-Musolf, S. Su and S. Tulin, {\it Phys. Rev}. {\bf D76}, 095017 (2007).

\bibitem{Neutrino}H. Lacker and A. Menzel, {\it JHEP} {\bf 07}, 006 (2010).

\bibitem{Leptoquarks}S. Davidson, D. Bailey and B. Campbell, {\it Z. Phys}. {\bf C61}, 613 (1994).

\bibitem{PRL}A. Aguilar-Arevalo {\it et al}., {\it Phys. Rev. Lett.} {\bf 115}, 071801 (2015) 

\bibitem{M13}A. Aguilar-Arevalo {\it et al}., {\it Nucl. Instrum. Methods.} {\bf A609}, 102 (2009).

\bibitem{NIMA}A. Aguilar-Arevalo {\it et al}., {\it Nucl. Instrum. Methods.} {\bf A791}, 38 (2015).

\bibitem{pimunug}G. Bressi {\it et al}., {\it Nucl. Phys}. {\bf B513}, 555 (1998).

\bibitem{NIM}A. Aguilar-Arevalo {\it et al}., {\it Nucl. Instrum. Methods.} {\bf A621}, 188 (2010).

\bibitem{LMLP}K.A. Olive {\it et al}., (Particle Data Group), {\it Chin. Phys}. {\bf C38}, 090001 (2014).

\end{thebibliography}
\end{document}